\documentclass[runningheads]{llncs}

\usepackage{svg}
\usepackage{graphicx}
\usepackage{soul}
\usepackage{graphicx}
\usepackage{amsmath}
\usepackage{booktabs}
\usepackage{url}
\usepackage{amssymb}
\usepackage{amsmath}
\usepackage{wrapfig}
\usepackage{multirow}
\usepackage{graphicx}
\usepackage{subfigure}
\usepackage{makecell}
\usepackage{algorithm}  
\usepackage{algorithmic}  
\usepackage{booktabs}
\begin{document}
\title{Empowering Sequential Recommendation from Collaborative Signals and Semantic Relatedness}
\author{Mingyue Cheng\inst{1}, Hao Zhang$^{1}$, Qi Liu \inst{1*}, Fajie Yuan\inst{2}, Zhi Li\inst{3}, Zhenya Huang\inst{1}, Enhong Chen\textsuperscript{1}, Jun Zhou\inst{4}, Longfei Li\inst{5} }
\institute{State Key Laboratory of Cognitive Intelligence, University of Science and Technology of China, Hefei, China \\ \email{\{mycheng, qiliuql, huangzhy, cheneh\}@ustc.edu.cn, zh2001@mail.ustc.edu.cn} \and Westlake University, Hangzhou, China \\ \email{yuanfajie@westlake.edu.cn} \and  Shenzhen International Graduate School, Tsinghua University, Shenzhen, China \email{zhilizl@sz.tsinghua.edu.cn} \and Ant Financial, Hangzhou, China  \\ \email{jun.zhoujun@antfin.com} \and Ant Financial Services Group, Hangzhou, China \\ \email{longyao.llf@antgroup.com}}
% \email{\{mycheng, qiliuql, huangzhy, cheneh\}@ustc.edu.cn,\{zh2001,yangjq, lili0516, wuli\_error, songliv\}@mail.ustc.edu.cn,zhilizl@sz.tsinghua.edu.cn}}

%% Use this when submitting an article to a sponsored event. You'll
%% receive a unique submission ID from the organizers
%% of the event, and this ID should be used as the parameter to this command.
%%\acmSubmissionID{123-A56-BU3}

%%
%% The majority of ACM publications use numbered citations and
%% references.  The command \citestyle{authoryear} switches to the
%% If you are preparing content for an event
%% sponsored by ACM SIGGRAPH, you must use the "author year" style of
%% the next command will enable that style.

%% end of the preamble, start of the body of the document source.

\maketitle
\let\thefootnote\relax\footnotetext{
% $^\dagger$Equal contribution, 
$^*$Qi Liu is corresponding author.}
\begin{abstract}
Sequential recommender systems (SRS) could capture dynamic user preferences by modeling historical behaviors ordered in time.
Despite effectiveness, 
focusing only on the \textit{collaborative signals} from behaviors does not fully grasp user interests.
It is also significant to model the \textit{semantic relatedness} reflected in content features, e.g., images and text. Towards that end, in this paper, we aim to enhance the SRS tasks by effectively unifying collaborative signals and semantic relatedness together. Notably, we empirically point out that it is nontrivial to achieve such a goal due to semantic gap issues. Thus, we propose an end-to-end two-stream architecture for sequential recommendation, named TSSR, to learn user preferences from ID-based and content-based sequence.  Specifically, we first present novel hierarchical contrasting module, including coarse user-grained and fine item-grained terms, to align the representations of inter-modality. Furthermore, we also design a two-stream architecture to learn the dependence of intra-modality sequence and the complex interactions of inter-modality sequence, which can yield more expressive capacity in understanding user interests. We conduct extensive experiments on five public datasets. The experimental results show that the TSSR could yield superior performance than competitive baselines. We also make our experimental codes publicly available at \url{https://github.com/Mingyue-Cheng/TSSR}. 
		
\keywords{Sequential Recommendation  \and Content Based Recommendation \and Contrastive Learning.}

\end{abstract}

	%\begin{itemize}
	%	\item {\texttt{acmconf}}: The default proceedings template style.
	%	\item{\texttt{sigchi}}: Used for SIGCHI conference articles.
	%	\item{\texttt{sigchi-a}}: Used for SIGCHI ``Extended Abstract'' articles.
	%	\item{\texttt{sigplan}}: Used for SIGPLAN conference articles.
	%\end{itemize}
	% \begin{figure}
	% 	\setlength{\belowcaptionskip}{-0.5cm}
	% 	\setlength{\abovecaptionskip}{0.1cm}
	% 	\centering
	% 	\includegraphics[width=1\linewidth]{figure_hico/1examples}

	% 	\caption{An illustrated of a user’s purchased behaviors.}

	% 	%	\Description{Block-wisely self-supervised neural architecture search methods.}
	% 	\label{fig:example1}
	% \end{figure}
\section{Introduction}
With the information overload, recommender systems~\cite{koren2008factorization,wu2022multi,zhao2021coea} have become one of the most useful information filtering engines for various online applications, such as e-commerce, advertising, and online videos. 
 % In the literature, collaborative filtering approaches~\cite{bai2017neural,koren2008factorization} could typically exhibit promising performances based on user-item interactions, becoming the most influential solutions for recommendation. 
 Among them, sequential recommender systems (SRS)~\cite{wang2019sequential,han2023guesr,wang2022transrec} have become  ever-increasingly prevalent due to their strengths in capturing dynamic user interests relative to traditional collaborative filtering approaches. 
 % The key setting of SRS is to learn high-quality user and item representations from each user's historically interacted behaviors ranked in chronological order.  
	Formerly, many efforts~\cite{hidasi2015session} have been devoted to designing sophisticated models to learn precise user and item embeddings in the SRS problem, including early traditional methods, e.g., matrix factorization combined with the Markov chain~\cite{rendle2010factorizing}, and recently various advanced neural models, including recurrent neural networks (RNNs), convolutional neural networks (CNNs), self-attentive models~\cite{kang2018self}.
 
 Though achieving success, the recommendation performance is unsatisfactory due to the sparse interactions of items. Luckily, various content features related to each item can be obtained in the SRS tasks~\cite{he2016vbpr}, such as each item's image appearance and textual descriptions. In general, visual contents play a significant role in influencing user decisions. 
 % Taking an example from clothing recommendation as shown in Figure~\ref{fig:example1},  each image typically contains rich appearance information, such as color and style, which largely reflect the appeal of a particular item to visiting users. As such, it deserved to additionally leverage content features to enhance the sequential recommendations.
	Toward integrating content features to empower the SRS task, researchers have proposed some attempts in previous works~\cite{hidasi2015session,zhang2019feature}. Nevertheless, we argue that these previous works only achieve limited performance gains in mining content features for sequential recommendation due to ignoring the semantic difference in embeddings of item IDs and content features. 
 The main reason behind item ID is that corresponding learned user and item embeddings could represent the implicit \textit{collaborative signals} among users, e.g., co-occurrence patterns. In contrast, content features typically contain very specific descriptions of an item. In real applications, it is very common to find similar items in terms of content features within a session. 
 With the existence of \textit{semantic relatedness}, these high-quality content feature can be utilized in performing recommendations even without leveraging item ID as shown in   Table~\ref{tab:id_content}. Nevertheless, these two aspects of embeddings reflect the user preference from different views. 
	
	Based on the above analysis, we set the goal of absorbing the capacity of \textit{collaborative signals} and \textit{semantic relatedness} together by treating item IDs and content features as two modalities.
  However, simply concatenating these features didn't improve performance in our tests as shown in Table~\ref{tab:id_content}, likely due to the challenge of aligning both modalities in a shared latent space, which restricts the system's effectiveness.
To illustrate this, we used t-SNE to visualize item embeddings from ID-only and content-only recommendation methods, as displayed in Figure~\ref{fig:gap}. The visualizations support our claim of a significant semantic gap between the embeddings of item IDs and content features. We assert that it's essential to model not just the sequence dependencies within each modality but also the dependencies across modalities to fully grasp user interests.
 % as illustrated in Figure~\ref{fig:example1}. 
	
	To that end, in this work, we propose an end-to-end two-stream architecture based on treating item IDs and content features as two modalities for sequential recommendations, refer to TSSR. The key idea behind the proposed model is a carefully designed hierarchical contrasting module for solving the issue of natural unaligned representations of item IDs and content features. Specifically, two levels of contrasting objectives, including coarse user-grained and fine item-grained terms, are proposed to align the representations of inter-modality. Besides, we introduce a two-stream encoder design, enabling not only sequence dependence on intra-modality but also item relationships of inter-modality to be well modeled, simultaneously. Due to comprehensively understanding user preferences via unifying \textit{collaborative signals} and \textit{semantic signals}, the TSSR naturally exhibits more accurate recommendations compared to previous approaches. We summarize the contributions as follows:
	\begin{itemize}
			\item  We highlight that unifying collaborative signals and semantic relatedness together could benefit the SRS task, in which the most distinctive characteristic is to group ID category identity and content features as two modalities. 
			\item We propose TSSR, an end-to-end two-stream architecture for sequential recommendation. To solve the key challenge, a novel hierarchical contrasting module is designed to alleviate the semantic gap across modalities. 
			\item We conduct extensive experiments on public datasets. The experimental results show that the TSSR exhibits superior recommendation results compared to previous competitive baselines. We also report some useful insights to show the influence of the TSSR in the SRS task. 
		\end{itemize}


	%~\cite{wang2022transrec}

	\section{Preliminaries}
% In this section, we first introduce the notations and formalize the sequential recommendation problem. Then, we perform empirical studies on benchmark recommenders over publicly available datasets. The novel contribution of this section is to uncover some findings to motivate our studies. 

\subsection{Problem Statement}
We introduce notations introduced in this paper and define the sequence recommendation problem. Assume that there are item set $\mathcal{I}=\{i_1, i_2, ..., i_{|\mathcal{I}|}\}$ and user set $\mathcal{U}= \{u_1, u_2, .., u_{|\mathcal{U}|}\}$, involving $|\mathcal{I}|$ items and $|\mathcal{U}|$ users. Assume that there are $|\mathcal{U}|$user item interaction sequence $\mathcal{X}= \{x_1, x_2, ..., x_{|\mathcal{U}|}\}$. Each sequence behavior contains user $u$'s historical interacted behaviors ranked in chronologically order, denoted by $\mathcal{X}_u = [x_1^u, x_2^u, ..., x_t^u]$, where $\mathcal{X}_u\in \mathcal{X}, u \in \mathcal{U}$. $x_l^u\in\mathcal{I}$ denotes the item user interacted at the time stamp $l$. $t$ is the sequence length. Assume that the content features of item $i$ are collected, like textual or visual descriptions of items, denoted by $c_i\in\mathcal{C}$, where $\mathcal{C}$ denotes the set of contents. Our goal is to predict the next item that users might interact with based on leveraging the multimodal sequence of item IDs and content features.
	\subsection{Empirical Studies}
	\label{sec:empirical}
	%To motivate our work, we perform extensive experiments over benchmark recommender models - SASRec proposed in~\cite{kang2018self}, whose effectiveness and efficient have been demonstrated in recent advancements. In addition to run the SASRec in the setting of ID-only sequences (dubbed as SASRec-ID), we further add a simple variants called SASRec-Content, involving content-only sequence for recommendation. The main reason behind such a setting is that we believe that embeddings of content features have been well encoded by these existing visual or textual representation models~\cite{he2016deep,devlin2018bert}. It should be noted that the compared methods strictly follow the hyper-parameter controlling rules. More implement details can be found in Section~\ref{sec:setup}. As shown in Table~\ref{tab:id_content}, we observe that the performance of content-only SRS could achieve comparable or more promising results compared to ID-only models. Such a observation largely break some settings in previous works~\cite{rendle2012bpr, he2016vbpr}, in which ID category identity information plays a dominating role in recommendations.
	We posit that content feature embeddings are already effectively captured by existing visual and textual models~\cite{he2016deep}, and it is common to find similar items within a session. Contrary to expectations, as Table~\ref{tab:id_content} reveals, content-only SRS performed comparably or even better than ID-only models. This suggests that the semantic relatedness within content feature sequences can significantly enhance recommendation quality. We believe this is reasonable because content embeddings encapsulate specific information, such as an item's style, size, and color.
		\begin{table}
		\centering
		\caption{Performance comparison(NDCG@10) on SASRec-ID and several variants i.e. SASRec-Content, SASRec-Hybrid.}
		% \box{1\linewidth}{!}
		\tabcolsep=0.3cm
		\begin{tabular}{cccccc}
			\toprule
			Method & H\&M  & Yelp  & Phone & Toy   & Mind \\
			% \midrule
			% SASRec-ID    & 0.1309 & 0.1058 & 0.0760 & 0.0731 & 0.2803 \\
			% SASRec-Content & 0.1515 & 0.1082 & 0.0830 & 0.0802 & 0.2922 \\
			% \hline
			% SASRec-Hybrid & 0.1506 & 0.0988 & 0.0828 & 0.0800  & 0.2895 \\
			\midrule
			SASRec-ID    & 0.0963 & 0.0463 & 0.0478 & 0.0478 & 0.1290 \\
			SASRec-Content & 0.1035 & 0.0491 & 0.0513 & 0.0503 & 0.1345 \\
			\hline
			SASRec-Hybrid & 0.1024 & 0.0425 & 0.0491 & 0.0506 & 0.1337 \\
			\bottomrule
		\end{tabular}
		% }%
	\label{tab:id_content}%
\end{table}%
		\begin{figure}[t]
		\centering
		\includegraphics[width=0.8\linewidth]{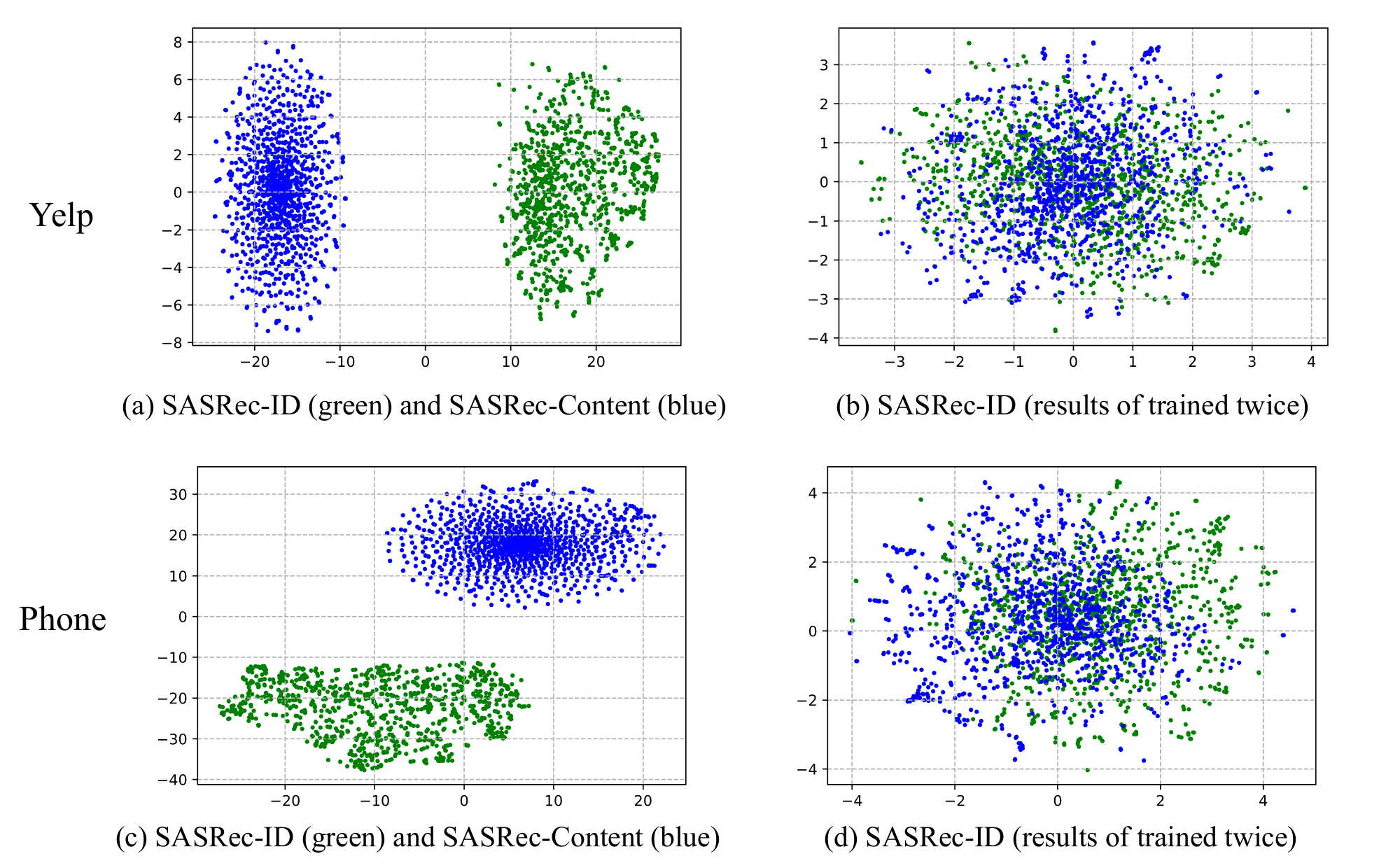}
		\caption{Visualization of item representations extracted from SASRec-ID and SASRec-Contents. It should be noted $1,000$ same items are shown in same line of figures, in which the up line denotes the results in the Yelp dataset while the below line reports the results in the Phone dataset.}
		%	\Description{Block-wisely self-supervised neural architecture search methods.}
		\label{fig:gap}
	\end{figure}
	Driven by these findings, we further construct more powerful recommender models by unifying the modalities of item IDs and content features together, enabling a more comprehensive understanding of user interests. Specifically, we construct a simple hybrid variant over SASRec, named SASRec-Hybrid, which concatenate two aspects of modality embeddings together before feeding to recommender models. As illustrated in Table~\ref{tab:id_content}, the experimental results show such a fusion seems cannot bring performance gains compared to unimodal input. Taking the difference of semantic embeddings of inter-modality, we further visualize the item representations extracted from several compared models as shown in Figure~\ref{fig:gap}. The visualization results largely demonstrate that there are big semantic gaps across modalities, suggesting that it is nontrivial to effectively fuse embeddings of item IDs and content features. %together in user interest modeling. 

		\begin{figure*}[t]
		\centering
		\includegraphics[width=\linewidth]{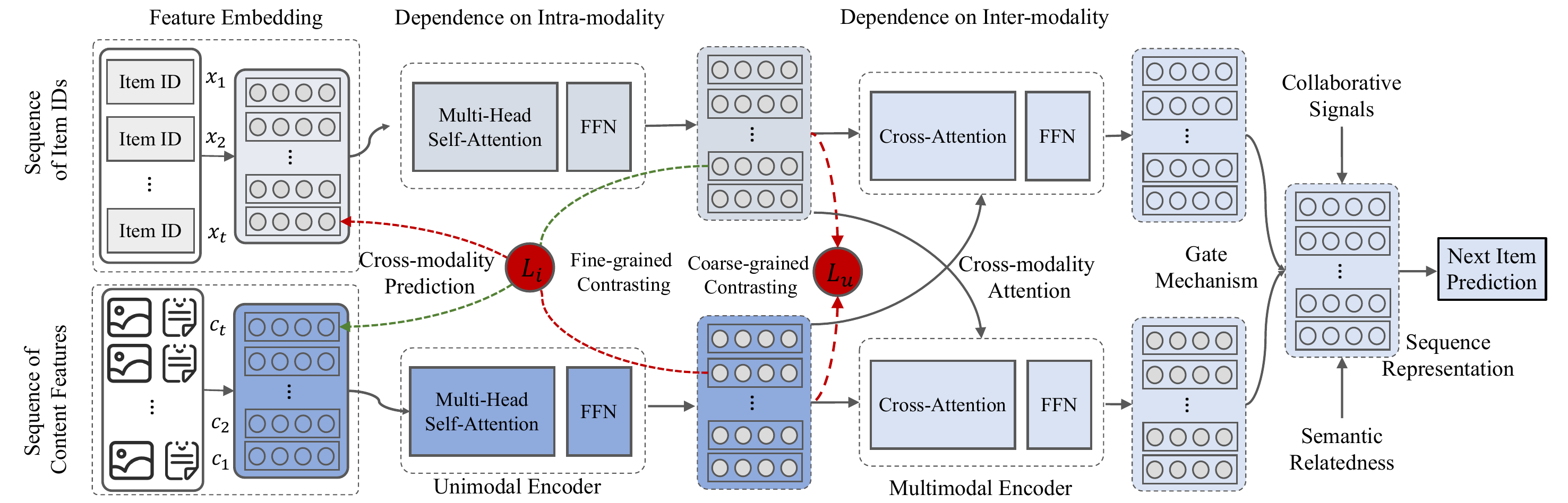}
		\caption{Illustration of the TSSR model, a two-stream architecture for performing sequential recommendation, in which collaborative signals and semantic relatedness are unified together.}
		%	\Description{Block-wisely self-supervised neural architecture search methods.}
		\label{fig:hico}
	\end{figure*}
	\section{The Proposed Model}
	Driven by these empirical studies above, we aim to present a focused study on unifying collaborative signals and semantic relatedness together so as to yield more accurate recommendations. One of the key challenges is to overcome the semantic gap issues across embeddings of item IDs and content features. In the next, we first present our designed TSSR, an end-to-end two-stream sequential recommendation model to capture sequence dependence from intra-modality and inter-modality sequences.  Next, we start by giving an overview of the proposed TSSR, followed by describing the model architecture in detail. 
	\subsection{Model Architecture Overview}
	The architecture of our proposed TSSR is depicted in Figure~\ref{fig:hico}. From left to right, four key layers are involved in the proposed TSSR architecture: feature representation layer, parallel unimodal encoder, parallel multimodal encoder, and user preference prediction. First, we employ an embedding layer to encode the sequence of item IDs and content features into latent spaces. Then, we adopt a parallel encoder design to capture the sequence dependence within modality and across modalities. Particularly, we develop a hierarchical contrasting module to align the representation of item IDs and content features. Lastly, these learned item representations are merged together with a simple gate mechanism. 
\subsection{Feature Representation}
Assume there are a sequence of interacted items denoted by $x = [x_1, x_2, ..., x_t]$ and the corresponding content sequences $c=[c_1, c_2, ..., c_t]$. Not loss of generality, we perform look-up operation from a learnable embedding matrix $\rm\textbf{M}^{id}\in\mathbb{R}^{|\mathcal{I}|\times d}$ to form the representations of item IDs where $d$ is the embedding dimension. As for the content feature representation, we hope our model can benefit from the previous pre-training works in language and visual domains.  Specifically, we regard the output of ResNet~\cite{he2016deep} as image appearance representations while employing the output of BERT~\cite{devlin2018bert} as the representations of textual descriptions. For each sequence, the corresponding ID-based and content-based representation can be denoted by $\rm\textbf{E}^{id}$ and $ \rm\textbf{E}^{con}\in\mathbb{R}^{t\times d}$. Note that these embeddings can be end-to-end fine-tuned during training.

\subsection{Aligning Unimodal Representations}
After obtaining the modality-specific representation of a given sequence, we then feed them into the encoder network so as to model the sequence dependence within modality just as pure ID-based SRS models.  In the TSSR, we decided to adopt the powerful transformer encoder designed in~\cite{kang2018self} to capture sequence dependence on intra-modality sequences. Its effectiveness and efficiency have been well verified in recent works~\cite{sun2019bert4rec}. And we also inject the corresponding positional encoding information, i.e., $\rm \textbf{P}^{id}$ and $\rm  \textbf{P}^{con}\in\mathbb{R}^{t\times d}$ to the corresponding representations to maintain the order of original sequence.
 % Though the self-attentive models can capture long-term dependence, self-attention itself is permutation-equivalent. Hence, we need to inject the corresponding positional encoding information, i.e., $\rm \textbf{P}^{id}$ and $\rm  \textbf{P}^{con}\in\mathbb{R}^{t\times d}$ to the corresponding representations of $\rm \textbf{M}^{id}$ and $\rm \textbf{M}^{con}$ to maintain the order of original sequence. After injecting position embedding, IDs-based and content-based embeddings are reformulated by $\rm \widehat{\textbf{E}}^{id}$ and $\rm \widehat{\textbf{E}}^{con}$. Here, we allow the positional embedding 
	% % matrix to be
	% learnable since its effectiveness has been verified in previous works~\cite{kang2018self,devlin2018bert}.  
	To model the sequence dependence within the unimodal sequence, two individual transformer encoders are utilized in parallel. After $L_{uni}$ layers non-linear transformation, representation of each sequence input is projected as $\rm \textbf{F}^{id}\in\mathbb{R}^{t\times d}$ and $\rm \textbf{F}^{con}\in\mathbb{R}^{t\times d}$. Although the network parameters of dual-encoder can be shared, we find such a setting cannot outperform the independent parameter setting. 
 The main reason is that there probably are big differences in transition patterns across modalities due to the nature of the data.
 
 % Actually, such dual-encoder architecture can be very flexible since we can fine-tune the individual networks w.r.t. their hyper-parameters. Besides, another important reason behind this type of architectural design is that there probably are big differences in transition patterns across modalities due to the nature of the data. The information density of content feature representation tends to be very detailed, whose corresponding representations can serve the whole recommender with semantic relatedness. Taking clothes as an example, image appearance can directly show the color, and text description can provide the clothing size. By contrast, the semantics of representation w.r.t discrete item IDs mainly encode the implicit collaborative signals of sequential behaviors. 
	
After exploring the item relationships within individual modalities, combining two modal representations poses significant challenges due to semantic gaps, as shown in Section~\ref{sec:empirical}. To address this, we introduce a hierarchical contrasting strategy that aligns the unimodal encoder outputs from two aspects: User-grained Contrasting and Item-grained Contrasting. 
	\paragraph{\textbf{User-grained Contrasting}}
To align two unimodal representations, we adopt a contrastive loss approach, proven effective in unsupervised learning breakthroughs~\cite{radford2021learning}. Item ID representations capture collaborative signals, while content features reflect semantic connections, both crucial for profiling user preferences. Therefore, we consider inter-modality embeddings as dual perspectives of the same user behavior, without relying on data augmentation. The calculation of the user-level contrastive loss $\mathcal{L}_u$
  is detailed as follows,
%	\begin{equation}
%		\begin{split}
%			\rm \mathcal{L}_{u} = -\dfrac{1}{|\mathcal{U}|}\sum_{u = 1}^{|\mathcal{U}|}( \log \dfrac{\exp(s(\widetilde{\textbf{F}}_u^{id}, \widetilde{\textbf{F}}_u^{con})/\tau)}{\sum_{j\in \mathcal{K}}\exp(s(\widetilde{\textbf{F}}_u^{id}, \widetilde{\textbf{F}}_j^{con})/\tau)} +\\ \rm \log \dfrac{\exp(s(\widetilde{\textbf{F}}_u^{con}, \widetilde{\textbf{F}}_u^{id}))/\tau)}{\sum_{j\in \mathcal{K}}\exp(s(\widetilde{\textbf{F}}_u^{con}, \widetilde{\textbf{F}}_j^{id})/\tau)}) ,
%		\end{split}
%		\label{equ:user}
	\begin{equation}
 \small
		\begin{split}
			 \mathcal{L}_{u} = -\dfrac{1}{|\mathcal{U}|}\sum_{u = 1}^{|\mathcal{U}|} \left( \log \dfrac{\exp(s(\widetilde{\textbf{F}}_u^{id}, \widetilde{\textbf{F}}_u^{con})/\tau)}{\sum_{j\in \mathcal{K}}\exp(s(\widetilde{\textbf{F}}_u^{id}, \widetilde{\textbf{F}}_j^{con})/\tau)} + \log \dfrac{\exp(s(\widetilde{\textbf{F}}_u^{con}, \widetilde{\textbf{F}}_u^{id}))/\tau)}{\sum_{j\in \mathcal{K}}\exp(s(\widetilde{\textbf{F}}_u^{con}, \widetilde{\textbf{F}}_j^{id})/\tau)} \right) ,
		\end{split}
		\label{equ:user}
	\end{equation}
\noindent where $\tau$ is the temperature used to scale the whole distribution.  $\mathcal{K}$ is an instance set, which contains a positive example and $|\mathcal{K}|-1$ negative ones. In addition, $s$ is the similarity measurement function. Here, we adopt the cosine to calculate the similarity of contrasted user representations by following~\cite{cheng2021learning}. $\rm \widetilde{\textbf{F}}^{id}$ and $\rm \widetilde{\textbf{F}}^{con}$ are used to denote the whole sequence representations w.r.t. item IDs and content features by performing meaning operation over all time steps of representation,  i.e., $\rm \textbf{F}^{id}$ and $\rm \textbf{F}^{con}$. 
	
	\paragraph{\textbf{Item-grained Contrasting}}
	While user-level contrastive loss efficiently aligns user sequence representations across modalities broadly,  it only remains the in a rough manner. Furthermore, we introduce a detailed cross-modality prediction task. Here, we predict the future steps of one modality $\rm x_{1:l}$ using the context $\rm c_{l+1}$ of another and vice versa.  Accordingly, we calculate the contrasting loss as: 
	\begin{equation}
 \small
		\begin{split}
			 \mathcal{L}_{i} = -\dfrac{1}{|\mathcal{U}|}\sum_{u = 1}^{|\mathcal{U}|} \sum_{l=1}^{t-1}\left ( \log \dfrac{\exp(s(\textbf{F}_{ul}^{id}, \textbf{E}_{u(l+1)}^{con})/\tau)}{\sum_{j\in \mathcal{G}}\exp(s(\textbf{F}_{ul}^{id}, \textbf{E}_j^{con})/\tau)} +  \log \dfrac{\exp(s(\textbf{F}_{ul}^{con}, \textbf{E}_{u (l+1)}^{id}))/\tau)}{\sum_{j\in \mathcal{G}}\exp(s(\textbf{F}_{ul}^{con}, \textbf{E}_j^{id})/\tau)} \right),
		\end{split}
		\label{equ:item}
	\end{equation}\noindent in which $\rm \textbf{F}_{ul}^{id}$ is the item representation of $l$-th time step in sequence $u$ while $\rm \textbf{E}_{u(l+1)}^{con}$ denotes the embeddings of item in the $(l+1)$-th time step extracted from embedding matrix $\rm \textbf{E}^{id}$, and vice versa for representations of $\rm \textbf{F}_{ul}^{con}$ and $\rm \textbf{E}_{u(l+1)}^{id}$. Here, we employ the inner product as the similarity measurement function $s$, whose effectiveness in terms of user preference prediction has been demonstrated in recent works~\cite{rendle2020neural}. In addition, the set $\mathcal{G}$ consists of a positive example and $|\mathcal{G}|-1$ negative instances. 
	In contrast to the original InfoNCE loss~\cite{oord2018representation}, such designed alignment loss in Eq~(\ref{equ:user}) and Eq~(\ref{equ:item}) can be treated as symmetric variants. Intuitively, the contrasting loss is still equal to optimize the lower bound on the mutual information between different modalities of the same historical behaviors.  In our designed hierarchical contrasting loss, the other positive examples within the same batch are simply treated as negative instances for optimization.  We are aware of more advanced negative sampling techniques might be helpful for further improving the results of the alignment module. We leave it for future extensions since it is not the focus of this work. 
 % In the experiments, we also detail the effects of the proportion of negative examples in experimental results.

	\subsection{Multimodal Interaction and Fusion}
	Building on the insights from recent studies~\cite{radford2021learning}, which highlight the advantages of leveraging multiple views of data for better object understanding, we aim to integrate multimodal information to enhance our comprehension of users' interests. To this end, we process the modality-specific representations through two separate multimodal encoders running in parallel. The unique aspect of these multimodal encoders is the substitution of self-attention with cross-attention mechanisms. This approach involves using the representation from one modality as the query and the representation from the other as the key and value in the attention calculation. This design's underlying principle is to enable each modality's representation to engage with and incorporate information from the other modality. After $\rm L_{multi}$ layers multimodal aggregating, the sequence representations of IDs and contents are denoted by $\rm \overline{\textbf{F}}^{id}, \overline{\textbf{F}}^{con}\in\mathbb{R}^{t\times d}$.
	
	To obtain the comprehensive item representations, we would merge the representation of two streams. Here, we devise a simple gate mechanism to learn the weight of each stream. After getting the output of sequence embedding, we project them into vectors by concatenation operation followed by a linear projection layer. Then, we compute the gate weight with the sigmoid function. Then each gating weight is attending to the corresponding stream’s output and is packed as the final feature vector. Formally, the whole fusion process can be described as,
	\begin{equation}
		\label{equ:gate}
		\begin{split}
			& \rm \textbf{S} = sigmoid (\textbf{W} [\overline{\textbf{F}}^{id},  \overline{\textbf{F}}^{con}] + b), \\
			& \rm \textbf{F} = (\textbf{S} \odot \overline{\textbf{F}}^{id}) + ( (\textbf{1}-\textbf{S}) \odot \overline{\textbf{F}}^{con}),
		\end{split}
	\end{equation}\noindent where $[\cdot]$ is the concatenation operation, and $\odot$ indicates the element-wise multiplication calculation. $\textbf{W}\in \mathbb{R}^{2d \times d}$ is a projection matrix and $b$ indicates the bias term, and $\textbf{S} \in \mathbb{R}^{t \times d}$ denotes gate matrix.

	\subsection{Model Optimization}
	%Formally, to model the joint distribution of item sequence $\mathcal{X}_u = [x_1^u, x_2^u, ..., x_t^u]$, we factorize it as a product of conditional distributions by the chain rule:
	%\begin{equation}
	%	p(i_{t+1}= i |i_{1:t}) = \prod_{i=2}^{t} p(x_i|x_{1:i-1}, \theta)p(x_1),
	%\end{equation}\noindent in which the $\theta$ denotes the model parameters of trained models.
	After obtaining the multimodal representation of whole sequence $\textbf{F}$, we next calculate each user's preference score for the item $i$ in the step $(t+1)$ under the input context with the inner product by following~\cite{kang2018self,cheng2021soft}. For the next item recommendation prediction task, we employ the auto-regressive generative loss as optimization objectives, i.e., left-to-right supervision signals, whose effectiveness has been demonstrated in previous works~\cite{cheng2022towards}. In this work, we employ cross-entropy loss combined with such optimization objectives (denoted by $\mathcal{L}_{\rm ce}$) to help train the whole model. Lastly, the proposed TSSR architecture can be optimized in a multi-task manner  with the trade-off parameters $\lambda_1, \lambda_2, \lambda_3 $ for each objective as follows:
	\begin{equation}
		\label{equ:training}
		\rm \mathcal{L} = \lambda_1 \cdot \mathcal{L}_u+  \lambda_2\cdot \mathcal{L}_i + \lambda_3 \cdot \mathcal{L}_{ce}, 
	\end{equation}

	\section{Experiments}
	% We first describe the experimental setup and compare them with other competitive baselines. We then analyze the experimental results of recommendation performance.
	\subsection{Experimental Setup}
	\label{sec:setup}
	\subsubsection{Datasets}
	We conduct experiments on five public datasets:
	\begin{itemize}
		\item \textbf{H\&M} is a publicly available dataset released in Kaggle Challenge 2022\footnote{https://www.kaggle.com/c/h-and-m-personalized-fashion-recommendations}. We take the historical purchase behaviors into account, in which each item's image appearance is regarded as the corresponding content features. And we randomly pick $100$ thousand users in experiments.
		\item \textbf{Yelp} is a public dataset~\footnote{https://www.yelp.com/dataset}, where we obtain each user's sequential behaviors by ranking their rated items in time order. The image appearance is treated as content features of items.  
		\item \textbf{Phone} and \textbf{Toy} are two datasets selected from Amazon review data\footnote{http://deepyeti.ucsd.edu/jianmo/amazon/index.html}, consists of ``Cell Phones and Accessories'' (Phone), ``Toys and Games'' (Toy), respectively. And we construct sequential behaviors by using the user's rating histories and regard the image of each item as the contents. 
		\item \textbf{MIND} is an English dataset for monolingual news recommendation\footnote{https://msnews.github.io/}. For each user, their history click logs are used to construct sequential behaviors, whose title textual descriptions are treated as contents.  
	\end{itemize}
	% Note that end-to-end training content-based feature encoder, e.g., BERT, is more computationally expensive  than pre-extracted features. Hence, 
 	\begin{table*}[t]
		\centering
		\caption{Statics of the datasets after processing.}
		\resizebox{1\columnwidth}{!}{
		\begin{tabular}{ccccccc}
			\hline
			Datasets & Num. Users & Num. Items & Num. Actions & Avg Actions/Item & Avg Actions/User & Content Modality \\
			\hline
			H\&M  & 100,000 & 61,574 & 692,413 & 11.2  & 6.9   & Image \\
			Yelp  & 173,318 & 36,208 & 1,348,192 & 37.2  & 7.8   & Image \\
			Phone & 128,647 & 42,675 & 841,657 & 19.7  & 6.5   & Image \\
			Toy & 141,303 & 55,285 & 996,576 & 18.0  & 7.1   & Image \\
			MIND  & 768,079 & 81,496 & 10,437,301 & 128.1  & 13.6  & Text \\
			\hline
		\end{tabular}}%
		\label{tab:dataset}%
	\end{table*}%
 We set the fixed sequence length of 20 for MIND while setting other datasets as $10$.  
	% It should be noted that if the sequence length is less than $t$, zero-padding can be used on the left side of the sequence to ensure the user's sequence to the same length. If the sequence is larger than $t$, recently $t$ items would be selected. 
	% For Yelp and Amazon, some shorter sequential behaviors, whose sequence length is less than $5$, and items that appear less than five times are filtered out.
	Besides, we filter items and users with fewer than five interaction records.
	 The statics of these datasets after processing are summarized in Table~\ref{tab:dataset}. 
	\subsubsection{Compared Methods}
	To verify the effectiveness of the TSSR, we choose representative methods as baselines, including traditional methods and deep sequential models. In particular, we regard some hybrid methods as baselines, in which both content and interaction records are simultaneously modeled. Note that we do not compare recent graph-based methods which are mainly studied for session-based recommendation. Next, we detail the compared baselines, 
	\begin{itemize}
		\item \textbf{PopRec} is a popularity-based method, in which each user is recommended according to the popularity distribution. 
		\item \textbf{ItemKNN}~\cite{sarwar2001item} is an item-based collaborative filtering method, where we employ an adaptive KNN neighbor as a recommendation algorithm.
		\item \textbf{BPR-MF}~\cite{rendle2012bpr} is a well-known matrix factorization-based methods optimized by Bayesian personalized ranking loss.
		\item \textbf{GRU4Rec}~\cite{hidasi2015session} is a pioneering attempts by employing recurrent neural network (RNNs) for next item recommendation. 
		%	\item \textbf{CL4Rec}~\cite{xie2022contrastive} utilizes contrastive learning to extract the discrimination Information in sequential recommendation.
		\item \textbf{MV-RNN}~\cite{cui2018mv} employs the RNNs as a sequence encoder to capture user interests from hybrids of item IDs and content features. \textbf{MV-TRM}~\cite{cui2018mv} is a variant of MV-RNN, in which we replace the RNNs architecture with a transformer encoder network for a fair comparison.  
		\item \textbf{CL4SRec}~\cite{xie2022contrastive} utilizes contrastive learning to extract the discrimination Information in sequential recommendation. 
		\item \textbf{FDSA}~\cite{zhang2019feature} first regards heterogeneous features of items into sequences with different weights with attention mechanisms, and integrates the outputs to make preference predictions. 
		\item \textbf{MFN4Rec}~\cite{wang2021auxiliary} treat the content feature as the side information by employing a multi-view gated memory network for sequential recommendation. 
		\item \textbf{SASRec-ID}~\cite{kang2018self} is a powerful SRS method based on self-attentive-based architectures to capture long-range dependence of sequential behaviors.  
		\item \textbf{SASRec-Content}~\cite{kang2018self} is a variant of SASRec, in which we directly regard the content feature representation as sequence input. 
	\end{itemize}
	% Noted that it preserves the same input between several content-enriched recommendation models and our proposed TSSR. 
 \begin{table*}[t]
		\centering
		\caption{Recommendation performance comparison of different recommendation models over five datasets, in which the best results are in bold and the ``IMP'' denotes the relative improvements our methods compared to SASRec-ID~\cite{kang2018self}.}
			\resizebox{1\columnwidth}{!}{
		\begin{tabular}{c|ccccc|ccccc}
			\hline
			\multirow{2}[2]{*}{Methods} & \multicolumn{5}{c|}{Recall@10}        & \multicolumn{5}{c}{NDCG@10} \\
			& H\&M  & Yelp  & Phone & Toy & MIND  & H\&M  & Yelp  & Phone & Toy & MIND \\
			\hline
			PopRec & 0.0105  & 0.0112  & 0.0076  & 0.0048  & 0.0220  & 0.0052  & 0.0059  & 0.0036  & 0.0025  & 0.0099  \\
			ItemKNN & 0.0456  & 0.0256  & 0.0097  & 0.0194  & 0.0138  & 0.0397  & 0.0122  & 0.0062  & 0.0117  & 0.0065  \\
			BPR   & 0.0180  & 0.0459  & 0.0206  & 0.0243  & 0.0151  & 0.0130  & 0.0237  & 0.0136  & 0.0146  & 0.0070  \\
			GRU4Rec & 0.0605  & 0.0661  & 0.0459  & 0.0421  & 0.2004  & 0.0416  & 0.0346  & 0.0318  & 0.0288  & 0.1100  \\
			CL4SRec & 0.1249  & 0.0668  & 0.0590  & 0.0635  & 0.1931  & 0.0991  & 0.0367  & 0.0416  & 0.0464  & 0.1059  \\
			FDSA  & 0.1104  & 0.0658  & 0.0549  & 0.0572  & 0.1940  & 0.0879  & 0.0348  & 0.0389  & 0.0425  & 0.1067  \\
			MFN4Rec & 0.0877  & 0.0655  & 0.0524  & 0.0520  & 0.1883  & 0.0603  & 0.0345  & 0.0350  & 0.0360  & 0.1040  \\
			MV-RNN & 0.0664  & 0.0597  & 0.0484  & 0.0408  & 0.1965  & 0.0416  & 0.0315  & 0.0327  & 0.0267  & 0.1078  \\
			MV-TRM & 0.1280  & 0.0628  & 0.0646  & 0.0657  & 0.2024  & 0.0967  & 0.0335  & 0.0445  & 0.0470  & 0.1118  \\
			SASRec-ID & 0.1179  & 0.0687  & 0.0593  & 0.0617  & 0.1949  & 0.0930  & 0.0364  & 0.0414  & 0.0450  & 0.1075  \\
			SASRec-Content & 0.1292  & 0.0697  & 0.0632  & 0.0654  & 0.2041  & 0.0978  & 0.0366  & 0.0428  & 0.0466  & 0.1123  \\
			TSSR  & \textbf{0.1419} & \textbf{0.0733} & \textbf{0.0684} & \textbf{0.0688} & \textbf{0.2107} & \textbf{0.1129} & \textbf{0.0387} & \textbf{0.0458} & \textbf{0.0485} & \textbf{0.1165} \\
			\hline
			IMP(\%) & 9.83\% & 5.16\% & 8.23\% & 5.20\% & 3.23\% & 15.44\% & 5.74\% & 7.01\% & 4.08\% & 3.74\% \\
			\hline
		\end{tabular}}%
		\label{tab:results}%
	\end{table*}%
	\subsubsection{Evaluation Metrics.}
	In this work, Recall and NDCG are employed as evaluation metrics to verify the effectiveness of compared methods. The former is an evaluation of un-ranked retrieval sets while the latter reflects the order of ranked lists. Here, we consider top-N (e.g., N = 10, 20) for recommendations. We follow the leave-one-out strategy to described in~\cite{kang2018self} to split the interactions into training, validation, and testing sets. For the testing stage, we adopt the all-ranking protocol to evaluate recommendation performance.
	\subsubsection{Hyper-parameter Settings}
	%We conduct all experiments on the environment of four 32GB NVIDIA V100. 
	For a fair comparison, all compared methods are trained until achieving the best recommendation results on the testing set. The embedding size is fixed at $128$ for all methods. We consistently employ Adam as the default optimizer, combined with a learning rate of $1\times e^{-4}$ and batch size of $256$. In all transformer encoder networks, the number of multi-head self-attention is $4$ while the number of FFN layers is $2$. For the remaining hyper-parameters and implement details in compared baselines, we either follow the default settings suggested by the original works or tune it on the validation set. For our proposed TSSR, the unimodal encoder block preserves the same depths as baselines while the multimodal encoder is set to $1$. The temperature in Eq~(\ref{equ:user}) and Eq~(\ref{equ:item}) loss is search from $\{0.1, 0.2,..., 1.0\}$. We set the trade-off of optimization objectives, including $\lambda_1, \lambda_2, \lambda_3$, as $1$.

	\subsection{Analysis of Experimental Results}

	\subsubsection{Recommendation Performance Evaluation}
The results in Table~\ref{tab:results} show that TSSR outperforms the competitive baselines, confirming the effectiveness of our approach in jointly modeling item ID and content feature sequences. Other content-rich SRS models fall short due to a lack of representation alignment or effective modeling of complex relations within or across modalities, resulting in sub-optimal performance. Additionally, we observe that TSSR particularly excels in the clothing domain, likely because visual elements are more influential in clothing selection than in other areas. However, improvements in scenarios with textual content are more modest, possibly because only news titles are considered as content. This suggests that developing methods for extracting textual representations is a promising avenue for future research.

 			\begin{figure}[t]
			\centering
	   \includegraphics[width=0.8\linewidth]{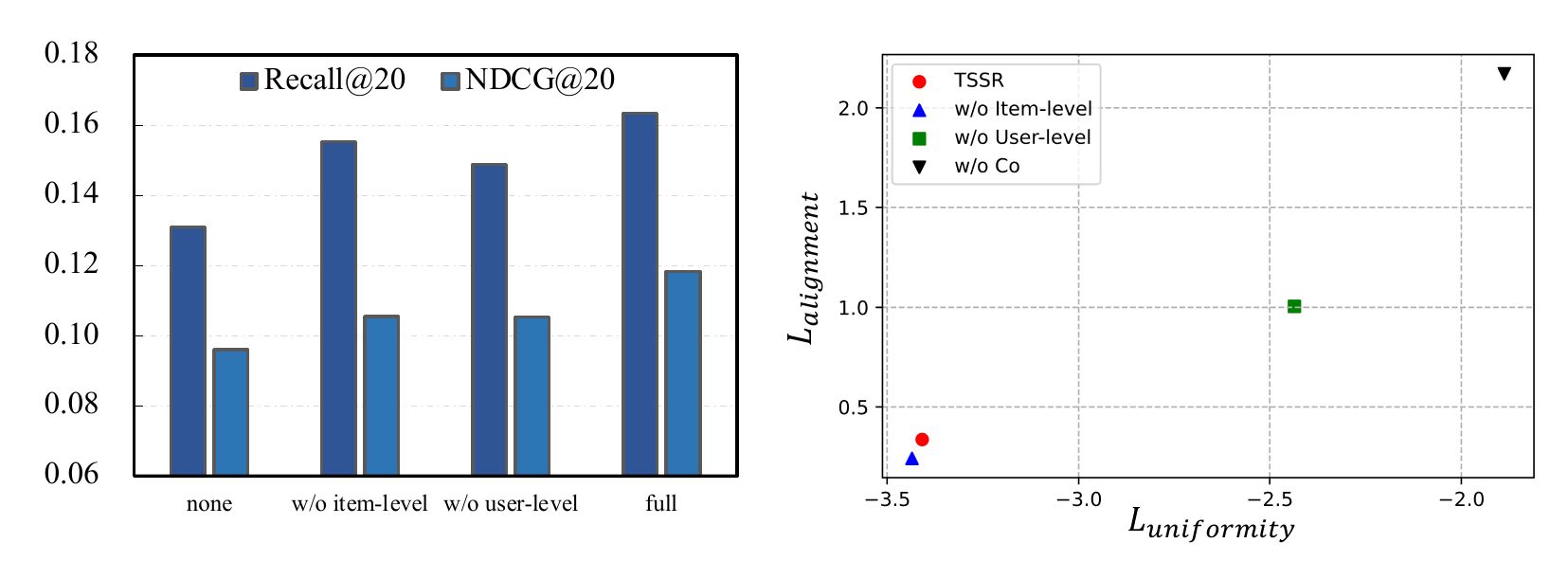}
			\caption{The effects of hierarchical contrasting for aligning the embeddings across different modalities, where left figure show the recommendation results and right figure reports the representation quality metric by alignment and uniformity.}
			\label{fig:ablation_contrasting}
		\end{figure}
        \begin{figure}[t]
        \hspace{4mm}
		\centering
		\includegraphics[width=0.8\linewidth]{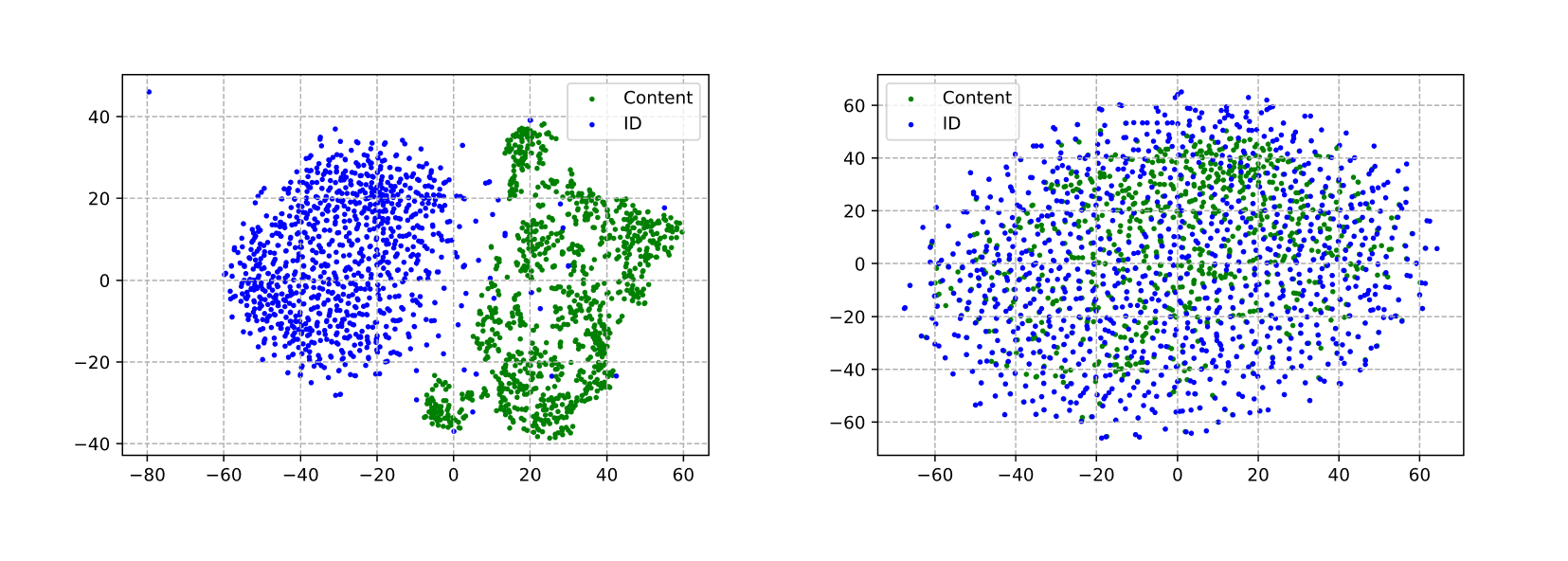}
  		\caption{Visualization of the t-SNE results of sampled items in the H\&M dataset, in which left figure denote the item representations without hierarchical contrasting while the right figure denotes the results our full model.}
		\label{fig:visualization_emb}
	\end{figure} 
	
	\begin{figure}[h]
		\centering
		\includegraphics[width=0.8\linewidth]{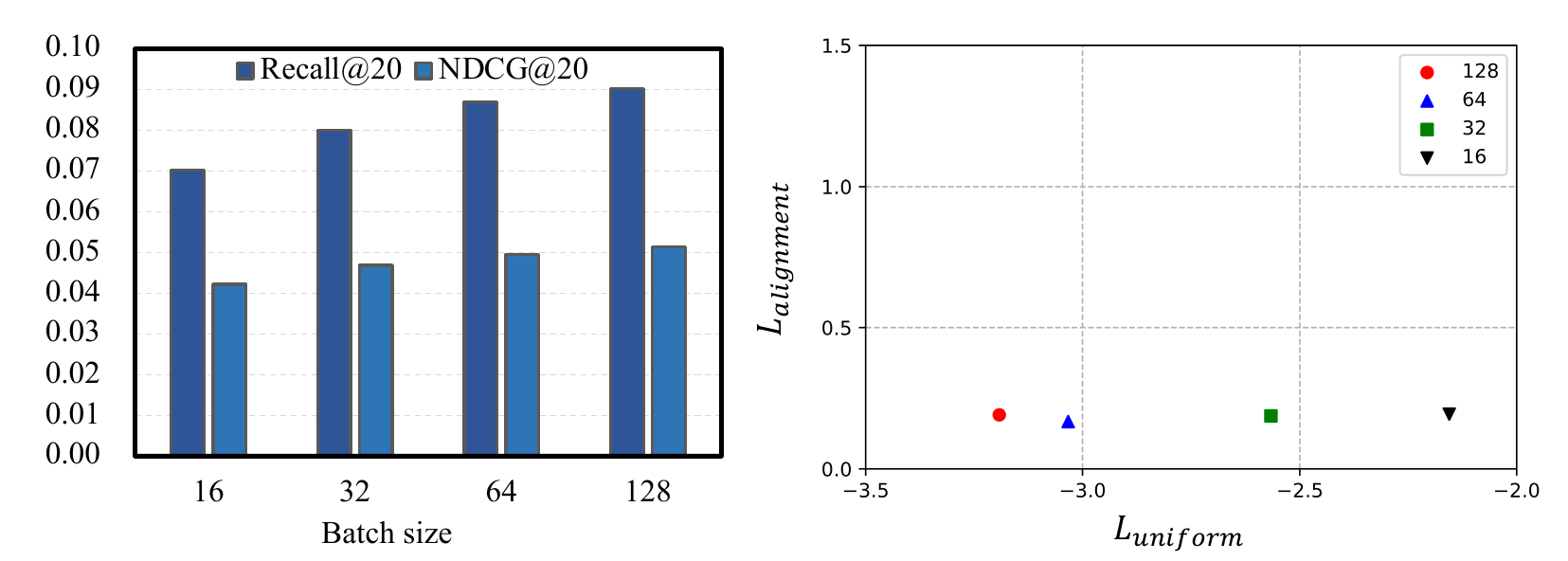}
		\caption{The impact of batch size in the TSSR model, where left figure report the recommendation performance and right figure shows the representation quality results measured by alignment and uniformity.}
		\label{fig:batch_size}
	\end{figure}
	
	\subsubsection{Study of Representation Alignment}
In this study, we delve deeper into the representation module of the full TSSR model by conducting ablation experiments. We test variants of the model where we remove the coarse user-level, fine item-level, or all alignment operations, respectively. We evaluate these variants on the H\&M dataset. Additionally, we assess the quality of the learned representations in terms of \textit{alignment} and \textit{uniformity} \cite{wang2020understanding}, which measure the closeness of paired instance embeddings and the even distribution of embeddings, respectively. It's important to note that we normalize all representations before evaluation.
Our findings, summarized in Figure~\ref{fig:ablation_contrasting}, show that the full representation alignment optimization indeed enhances the TSSR model. Interestingly, we observe that better alignment and uniformity in representations do not necessarily lead to improved recommendation performance and may even hinder it. We attribute these results to two main factors: first, the similarity between item-grained contrasting and generative auto-regressive objectives; second, the intrinsic nature of item recommendation as a ranking rather than a classification problem ~\cite{gao2021simcse}.
	To further intuitively show how well the item embeddings are aligned by our hierarchical contrasting module, we show the fusion performance of learned item embeddings by performing t-SNE over-extracted embeddings. As shown in Figure~\ref{fig:visualization_emb}, the left figure denotes the t-SNE visualization comparison across embeddings of item IDs and content features while the right figure indicates the visualization of aligned embeddings in the TSSR model. Such results further demonstrate the effectiveness of our TSSR in aligning representations of item IDs and content features.

	\subsubsection{Hyper-parameter Sensitivity Analysis w.r.t. Contrasting Optimization}
	To solve the semantic gap issue, we carefully develop the hierarchical contrasting module, in which the number of negative sampling plays a vital role in the final performance. Due to the mini-batch sharing strategy in our work, we hence turn to empirically evaluate how would batch size affects performance. For clarity and saving space, we only report the performance of TSSR trained in the Phone dataset as shown in Figure~\ref{fig:batch_size}. Note that we also examine the representation quality of learned embedding with \textit{alignment} and \textit{uniformity}. From the experimental results, we find that: 1) our proposed TSSR can benefit from a larger batch size, and 2) the batch size can also greatly influence the value of uniformity. We hold such an experimental phenomenon is reasonable since that larger batch size can provide more negative examples for hierarchical contrasting optimization. Hence, the representation quality of item representations is naturally improved with the increasing of negative examples. 

	\subsubsection{Result Analysis w.r.t. the Amount of Training Data }
 	\begin{figure}[t]
		\centering
	\includegraphics[width=0.8\linewidth]{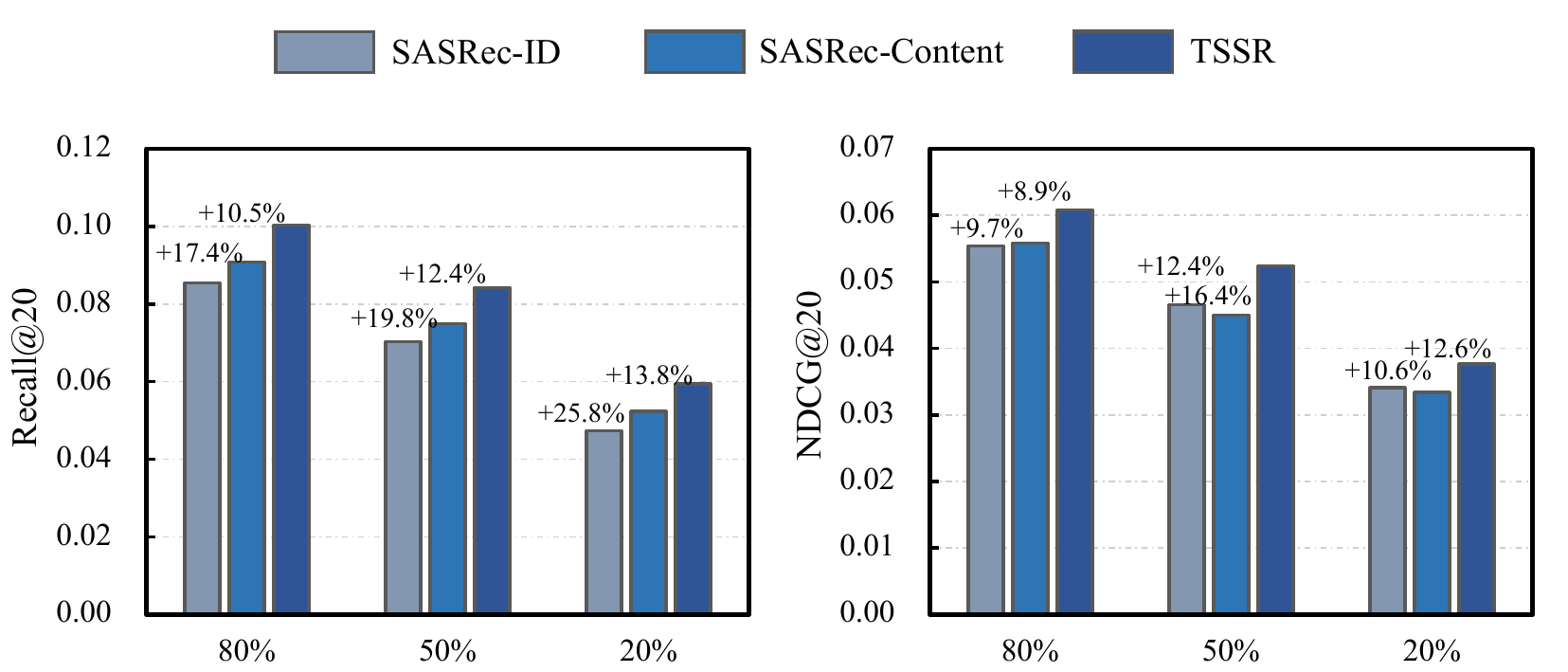}
		\caption{Recommendation performance comparison w.r.t. the amount proportion of training set in the Toy dataset.}
		\label{fig:cold}
	\end{figure}
	As one type of typical collaborative filtering, one of the unique challenges is easily suffering from the data sparse issue in real-world applications. Luckily, such a dilemma can be largely alleviated by the proposed TSSR since the newly designed hierarchical contrasting module actually improve the utilization of data by constructing self-supervised training signals. We hence simulate the data sparsity scenarios by using different proportions of the full training sets, i.e., 80\%, 50\%, 20\%.  Correspondingly, Figure~\ref{fig:cold} shows the evaluation results on the Toy datasets. As we can see, the performance substantially drops when a lower proportion of training data is used. However, our TSSR can consistently outperform baselines in all cases. Particularly, the TSSR could achieve superior relative improvements in an extreme sparsity situation (20\%). This observation implies that TSSR is able to make better use of the data by constructing self-supervised training signals, thus enabling our model more robust for sparse interaction scenarios. 
	\subsubsection{Limitations Analysis}
	Although our work can effectively organize item IDs and content features in a unified framework, we also notice 
the key limitation about high training cost of TSSR because of the end-to-end learning paradigm. For example, TSSR typically requires around 120G GPU memory in the H\&M dataset. In contrast, around 2G GPU memory is enough for ID-only baselines (i.e., SASRec-ID).  Luckily, we notice some adapter-based parameter-efficient learning techniques have been extensively studied, in which a very small proportion of trainable model parameters (e.g., only $3\%$ in the classic Houlsby adapter~\cite{houlsby2019parameter}) is able to achieve comparable performances. 
 % A second limitation is that we only examine TSSR with one type of content feature for each item. We are aware of aggregating multiple categories of content features into new feature sequences might benefit the final sequential recommendation performance. In future work, we plan to evaluate the effectiveness of our work by collecting multiple types of modality features of items. 

	\section{Related Work}
	% In this section, we briefly introduce the related work about sequential recommendation and content-enriched recommendation. 
	\paragraph{\textbf{Sequential Recommendation}}
	% \section{}{Sequential Recommendation}
	Sequential recommender systems (SRS)~\cite{wang2019sequential} have become ever-increasing prevalent by learning precise user and item embeddings from sequential behaviors. Formally, numerous efforts have been devoted to designing sophisticated models for the SRS scenario. A well-known traditional work is the FPMC model, which combines the Markov chain and matrix factorization techniques for the next basket recommendation. Later, researchers largely facilitate this area with various advanced neural model techniques.  As one pioneering attempt, GRU4Rec has been widely used to solve sequential recommendations by employing the hidden state of recurrent neural networks (RNNs) to propagate user preference. Due to its effectiveness, a series of extensions~\cite{li2017neural} have been proposed based on such models.  In addition, self-attentive and convolutional-based models can be very effective in modeling sequential behaviors due to their strengths in the capacity of long-term dependence modeling and easy parallel computation. 
\vspace{-0.1in}
	\paragraph{\textbf{Content-enriched Recommendation}}
	% \subsection{Content-enriched Recommendation}
	%Although collaborative filtering can be effective for numerous user-item interactions,  the learned embeddings may not be effective due to the sparse interaction scenarios. Hence,
	Content-enriched recommendations~\cite{wu2021survey} become very prevalent in alleviating the cold-start recommendation issue. Accompanied by the prevalence of deep learning, content features can be effectively embedded in latent space. In recommendation, visual-based multi-media contents,  including images and videos, and textual-based contents, like item descriptions or review texts have been extensively on general recommenders~\cite{he2016vbpr}. However, these works only consider the static preference of each user while ignoring the dynamic evolving interests. Hence, a series of works were also proposed by following the intuition of content information could greatly influence user's behavioral decision~\cite{hidasi2016parallel}.  Several previous works were also proposed by leveraging 3D convolution networks~\cite{tuan20173d}, self-attention mechanism~\cite{zhang2019feature}, memory networks~\cite{wang2021auxiliary}. Besides, some researchers also proposed to leverage the attribute information in terms of timestamp information~\cite{li2020time}, and abundant both user and item-side information~\cite{rashed2022carca}. 
		In addition, we also notice that some prior works have adopted contrastive learning techniques~\cite{xie2022contrastive} to benefit the representation user preference modeling. However, differently, we mainly present a focused study on unifying collaborative signals and semantic relatedness together by leveraging a novel contrasting optimization to solve the representation gap issue. 
	
	%In recommendation, visual-based multimedia content, e.g., images and videos, are the most eye-catching contents for users and can be very common on social media platforms. Related works were fully proposed based on mining fine-grained visual attributes, and domain properties et al. Considering that there is also textual information, such as reviews and textual descriptions, some autoencoder-based models, and word embeddings have been designed. 
	%As for sequential recommendation scenarios~\cite{rashed2022carca},
\vspace{-0.2in}
 \section{Conclusion}
	In this work, we proposed to effectively leverage collaborative signals and semantic relatedness together to improve the SRS task. To achieve this goal, we proposed an end-to-end two-stream architecture, named TSSR, whose most distinctive characteristic is to treat the ID identity and content features of items as two modalities. In the TSSR, the newly designed hierarchical contrasting module could effectively alleviate the significant challenges of semantic gap issues in fusing these two modality sequences in a unified architecture. Besides, the TSSR model could capture the sequence dependence within and cross modalities, simultaneously. We conduct extensive experiments on public datasets. The experimental results demonstrate that the TSSR could obtain a higher performance gain in jointly modeling item IDs and content features.
 \\\\
  \textbf{Acknowledgement} This research was supported by grants from the National Key Research and Development Program of China (Grant No. 2021YFF0901003) and the Fundamental Research Funds for the Central Universities.
%   \textbf{Acknowledgement} This research was partially supported by grant from the
% National Natural Science Foundation of China (Grant No. 61922073). This work
% also thanks the support of fundings MAI2022C007 and WK5290000003.
	% We hope our work could inspire more work to be proposed for multi-modal sequential recommendation.
	%In addition, we also report some useful insights to deeply understand this reformulated research goal and the newly designed model. 
	%We would like to extend our work to transferring recommendation scenarios.  
			% \newpage
\begin{small}
	\bibliographystyle{splncs04}
	\bibliography{tssr-sample-base}
\end{small}
	%\appendix
	%Lorem ipsum dolor sit amet, consectetur adipiscing elit. Morbi
	%malesuada, quam in pulvinar varius, metus nunc fermentum urna, id
	%sollicitudin purus odio sit amet enim. Aliquam ullamcorper eu ipsum
	%vel mollis. Curabitur quis dictum nisl. Phasellus vel semper risus, et
	%lacinia dolor. Integer ultricies commodo sem nec semper.
	%Etiam commodo feugiat nisl pulvinar pellentesque. Etiam auctor sodales
	%ligula, non varius nibh pulvinar semper. Suspendisse nec lectus non
	%ipsum convallis congue hendrerit vitae sapien. Donec at laoreet
	%eros. Vivamus non purus placerat, scelerisque diam eu, cursus
	%ante. Etiam aliquam tortor auctor efficitur mattis.
	
	%Nam id fermentum dui. Suspendisse sagittis tortor a nulla mollis, in
	%pulvinar ex pretium. Sed interdum orci quis metus euismod, et sagittis
	%enim maximus. Vestibulum gravida massa ut felis suscipit
	%congue. Quisque mattis elit a risus ultrices commodo venenatis eget
	%dui. Etiam sagittis eleifend elementum.
	%
	%Nam interdum magna at lectus dignissim, ac dignissim lorem
	%rhoncus. Maecenas eu arcu ac neque placerat aliquam. Nunc pulvinar
	%massa et mattis lacinia.
\end{document}